\documentclass[aricle,preprint,superscriptaddress,floatfix, 
nolongbibliography]{revtex4-2}
\usepackage[T2A]{fontenc}
\usepackage{graphicx}
\usepackage{amsmath}
\usepackage{amssymb}
\usepackage{siunitx}
\usepackage{cleveref}
\usepackage[english]{babel}
\usepackage[utf8]{inputenc}
\usepackage[T1]{fontenc}
\newcommand{\tauq}{\tau_q}

\newcommand{\omegac}{\omega_c}

\newcommand{\EF}{E_F}

\begin{document}

\title{Giant Shubnikov-de Haas Oscillations with V-Shaped Minima in a High-Mobility Two-Dimensional Electron Gas: Experiment and Phenomenological Model}

\author{E. Yu. Zhdanov}
\email{zhdanov@isp.nsc.ru}
\affiliation{Rzhanov Institute of Semiconductor Physics SB RAS, Novosibirsk 630090, Russia}
\affiliation{Novosibirsk State University, Novosibirsk 630090, Russia}

\author{M. V. Budantsev}
\affiliation{Rzhanov Institute of Semiconductor Physics SB RAS, Novosibirsk 630090, Russia}
\affiliation{Novosibirsk State University, Novosibirsk 630090, Russia}

\author{D. I. Sarypov}
\affiliation{Rzhanov Institute of Semiconductor Physics SB RAS, Novosibirsk 630090, Russia}
\affiliation{Novosibirsk State University, Novosibirsk 630090, Russia}

\author{D. A. Pokhabov}
\affiliation{Rzhanov Institute of Semiconductor Physics SB RAS, Novosibirsk 630090, Russia}
\affiliation{Novosibirsk State University, Novosibirsk 630090, Russia}

\author{A. K. Bakarov}
\affiliation{Rzhanov Institute of Semiconductor Physics SB RAS, Novosibirsk 630090, Russia}
\affiliation{Novosibirsk State University, Novosibirsk 630090, Russia}

\author{A. G. Pogosov}
\affiliation{Rzhanov Institute of Semiconductor Physics SB RAS, Novosibirsk 630090, Russia}
\affiliation{Novosibirsk State University, Novosibirsk 630090, Russia}

\date{\today}

\begin{abstract}
Giant Shubnikov-de Haas oscillations (SdHO) with V-shaped minima are experimentally studied in a high-mobility two-dimensional electron gas based on GaAs/AlGaAs heterostructures. A phenomenological model with two parameters (transport momentum relaxation time $\tau_{\text{tr}}$ and quantum scattering time $\tau_q$) is developed, accurately describing experimentally measured magnetoresistance over an unexpectedly wide range of magnetic fields (up to \SI{3.5}{\tesla}) and temperatures (from 2~K to 15~K). The model combines: (i) a quasiclassical density of states with a magnetic-field-dependent Gaussian broadening of Landau levels, (ii) a momentum relaxation time scaling with the density of states, and (iii) oscillations of the Fermi level at a fixed electron density. This model reproduces V-shaped oscillation minima with zero-resistance points, a smooth background of positive magnetoresistance, and enables the extraction of $\tau_q$ and $\tau_{\text{tr}}$ even in microstructures where ballistic and viscous effects dominate at low fields. 
As expected, the temperature dependence reveals that $\tau_{\text{tr}}$ scales inversely with temperature due to acoustic phonon scattering, while $\tau_q$ remains temperature-independent.

\end{abstract}

\maketitle

\section{Introduction}
Two-dimensional electron systems are a model system for studying collective and quantum effects at low temperatures and high magnetic fields. These systems exhibit the integer and fractional quantum Hall effects (IQHE, FQHE) \cite{Klitzing1980,Prange2012,Budantsev2014,Qian2017,Klitzing2020}, spin-related phenomena \cite{Knap2004,Piot2005,Minkov2016,Pokhabov2018,Myronov2023}, viscous electron flow \cite{Bandurin2016,Gusev2018,Gupta2021,Wang2023,Egorov2024,Sarypov2025,Sarypov2025slip}, and Shubnikov-de Haas oscillations (SdHO) \cite{Ando1982,Laikhtman1994,Endo2008,Monteverde2010,Zawadzki2013,Hang2009,Hayne1997,Hatke2012,Kunc2015,Peters2016,Mayer2016,Munasinghe2018,Huber2023}. SdHO remain one of the most sensitive methods for two-dimensional electron gas (2DEG) spectroscopy, as their amplitude and shape directly depend on the Landau level broadening $\Gamma$, governed by the quantum scattering time $\tau_q$ and disorder type.

Theoretical descriptions of SdHO are well-established only in specific magnetic field regimes. In weak fields, where the disorder correlation length is much smaller than the magnetic length ($l_{\text{corr}} \ll l_B = \sqrt{\hbar/(eB)}$), with strong Landau level overlap, the exponentially small oscillation amplitude follows the standard self-consistent Born approximation (SCBA) model \cite{Ando1982,Coleridge1989}, which yields a  Lorentzian broadening of the Landau levels. In moderate fields ($l_{\text{corr}} \sim l_B$), the modified SCBA model \cite{Vavilov2004} provides the most accurate description, according to Refs.  \cite{Tan2005,Dietrich2012}, incorporating correction factors into the density of states to match experimental SdHO profiles. Finally, in strong magnetic fields ($\omega_c\tau_q \gg 1$, $l_{\text{corr}} \gg l_B$) where Landau levels are well-separated, only the quasiclassical approach \cite{Raikh1993} is applicable. It leads to a Gaussian broadening of the Landau levels and accurately describes the density of states. However, this approach has not been used directly to calculate the amplitude and the shape of the magnetoresistance curve $R(B)$.

Experimental SdHO analysis typically relies on Dingle plots \cite{Coleridge1989,Coleridge1996,Dmitriev2012}, which allows accurate estimation of $\tau_q$ in weak fields ($\omega_c\tau_q \ll 1$). However, oscillations are often thermally suppressed here, and magnetoresistance in microstructured samples is modulated by ballistic and viscous effects \cite{Alekseev2016}. Standard Dingle analysis is insensitive to the Landau levels broadening models since all three aforementioned models are indistinguishable at $\omega_c\tau_q \ll 1$.

In this work, we experimentally demonstrate that existing models—including the modified SCBA model \cite{Vavilov2004} fail to describe details of the 2DEG magnetoresistance in high-density, high-mobility samples with weak disorder potential. The observed details of the magnetoresistance, such as shape, amplitude, asymmetry of oscillations, and positive background, deviate significantly from theoretical predictions. To address this, We propose an alternative phenomenological approach. While Ref. \cite{Endo2008} demonstrated that Gaussian broadening of Landau levels combined with Fermi level oscillations yields V-shaped minima in the density of states, their analysis stopped short of calculating the magnetoresistance $R(B)$ itself. In contrast, our phenomenological model extends this approach by self-consistently computing the magnetotransport via Eqs. \eqref{eq:resistance} - \eqref{eq:tauB} (see Sec. IV), enabling direct quantitative comparison with experimental $R(B,T)$ over a wide range of fields and temperatures. Our model combines three key ingredients: (i) Quasiclassical density of states with magnetic-field-dependent Gaussian Landau level broadening \cite{Raikh1993}; (ii) Momentum relaxation time $\tau_{\text{tr}}=(m/ne^2) \sigma_0$, where $\sigma_0$ is the conductivity of intrinsic unstructured 2DEG, scaling with density of states $\nu(B,E)$: $\tau_\text{tr} \to \tau_B(E) = \tau_\text{tr} \frac{\nu_0}{\nu(B,E)}$, where $\nu_0$ is zero-field value of the density of states at the Fermi level \cite{Coleridge1989,Dmitriev2012,Dmitriev2003}; and (iii) Magnetic-field-dependent oscillations of the Fermi level $E_F$ at a fixed electron density.

This combination allows us to reproduce not only the V-shaped minima but also the absolute amplitude and temperature evolution of the oscillations, and to extract the scattering times $\tau_{\mathrm{tr}}$ and $\tau_q$ with high accuracy, even in nanostructures where ballistic and hydrodynamic effects obscure the intrinsic 2DEG response. Moreover, the same unified model self-consistently accounts for the smooth positive magnetoresistance background observed at higher temperatures where SdHO are suppressed. We confirm that the scaling $1/\tau_B(E) \propto \nu(E)$ holds even under strong quantization, despite limited theoretical justification, and the known approach that $E_F$ oscillates with $B$, contrary to the widely spread assumptions of its constancy.

The approach is particularly relevant for micro/nanostructures where ballistic/hydrodynamic effects enhance resistance at $B \rightarrow 0$ \cite{Alekseev2016,Gupta2021,Pogosov2022}, preventing direct extraction of the 2DEG's intrinsic $\tau_{\text{tr}}$. For certain systems (e.g., suspended 2DEGs \cite{Pogosov2012,Pogosov2022,Zhdanov2017}), proposed approach is the only practical way to determine $\tau_{\text{tr}}$.

\section{Samples and Methods}
\begin{figure}[htbp]
\centering
\includegraphics[scale=1.25]{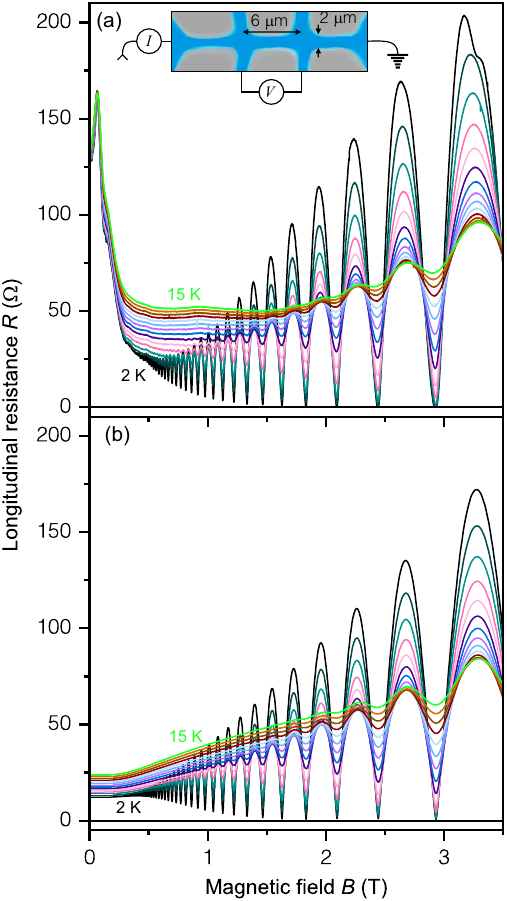}
\caption{
Experimental (a) and calculated (b) longitudinal magnetoresistance $R(B)$ at temperatures from 2~K to 15~K. Inset: Optical micrograph of the sample.
\label{fig1}
}
\end{figure}

Longitudinal magnetoresistance $R(B)$ of microscopic Hall bars was measured in magnetic fields up to \SI{3.5}{\tesla} and temperatures from 2~K to 15~K using an Oxford Instruments Teslatron PT system. A four-probe lock-in technique (\SI{70}{\hertz}, \SI{0.1}{\micro\ampere}) ensured linear response, with voltage detected synchronously (see the inset in Fig.~\ref{fig1}).

Samples were fabricated from AlGaAs/GaAs heterostructures grown by molecular beam epitaxy. The 2DEG resided in a \SI{13}{\nano\meter} quantum well flanked by AlAs/GaAs superlattices with Si $\delta$-doping layers ($d = \SI{30}{\nano\meter}$ on each side). It is expected that dual-side doping enhaces $\tau_\text{tr}$ by $\sqrt{2}$ \cite{Dmitriev2012}. Nonconducting X-valley electrons near $\delta$-layers enhanced mobility, as in Ref.~\cite{Friedland1996}. Key parameters: electron density $n = 7.08 \times 10^{11}$ \si{\per\square\centi\meter} ($E_F = \SI{290}{\kelvin}$), mobility $\mu = 2 \times 10^6$ \si{\square\centi\meter\per\volt\per\second}, and reduced $d$ compared to conventional high-mobility structures. Hall bars with width of $W = \SI{2}{\micro\meter}$ and length of $L = \SI{6}{\micro\meter}$ had aspect ratio $L/W =$3.

\section{Experimental Results}
Figure~\ref{fig1}a shows the magnetoresistance $R(B)$ at temperatures from 2~K to 15~K. SdHO with round-shaped maxima and V-shaped minima appear between 0.5~T and 3.5~T. At high magnetic fields and states with odd filling factors $\gamma$ and unresolved spin, resistance maxima exceed the $R(B = \SI{0.5}{\tesla})$ value by an order of magnitude. For even $\gamma$, the resistance minima approach zero above $B \sim \SI{1.5}{\tesla}$, but without QHE plateau-like zeroings, indicating negligible number of localized states for Fermi level pinning. In addition, the non-local resistance $R_{\text{nonlocal}}$ measured at $B \sim \SI{3}{\tesla}$ is not the same order of magnitude as $R$ as one would expect in the QHE regime, but is 100 times smaller than. Thus, the studied 2DEG does not reach the QHE regime at least up to 3.5 T.

Rise of temperature suppresses oscillations, reducing amplitude and shifting onset to higher fields. At \SI{15}{\kelvin}, low-amplitude oscillations start above \SI{1.5}{\tesla}, with a near-cosinusoidal form, as in the case of ordinary OSdH. The magnetoresistance background increases monotonically with temperature (similar behavior was also observed e.g. in Ref. \cite{Shi2014}). 
Notably, at low magnetic fields (up to \SI{0.3}{\tesla}), we observe a magnetoresistance peak, originating from ballistic electron scattering at rough boundaries \cite{Shi2014,Roukes1987,Budantsev1996,Zhdanov2017}. This behavior is characteristic of all high-quality microscopic samples. The ballistic peak is sharply suppressed by weak magnetic fields, when the cyclotron diameter becomes smaller than the sample width. At higher magnetic fields, Shubnikov–de Haas oscillations develop on top of a smoothly growing, field-dependent background. The substantial magnitude of the ballistic peak precludes the accurate determination of the zero-field 2DEG resistance $R_0 = \frac{L}{W}  \frac{m}{n e^2 \tau_{\text{tr}}}$, which would characterize a macroscopic sample, and consequently prevents the reliable extraction of $\tau_{\text{tr}}$. While the Dingle plot analysis could potentially resolve this limitation, this method does not consistently have acceptable accuracy, as will be shown in Sec.~\ref{sec:dingle}. Nevertheless, we introduce a model that not only provides both qualitative and quantitative descriptions of the experimentally observed magnetoresistance behavior (compare Figs.~\ref{fig1}a and \ref{fig1}b) across a wide magnetic field range, but also enables the extraction of the parameters $\tau_q$ and $\tau_{\text{tr}}$.

\section{Theoretical Model}
\subsection{Magnetoresistance Formalism}
Longitudinal resistance is calculated as:
\begin{equation}
R = \frac{L}{W} \frac{\sigma_{xx}}{\sigma_{xx}^2 + \sigma_{xy}^2},
\label{eq:resistance}
\end{equation}
with conductivity tensor components:

\begin{equation}
\sigma_{xx} = \sigma_0 \int_0^\infty dE \left( -\frac{\partial f}{\partial E} \right) \frac{1}{1 + \omega_c^2 \tau_B^2(E)} ,
\label{eq:sigmaxx}
\end{equation}

\begin{equation}
\begin{split}
    \sigma_{xy} = -\frac{e n}{B} +  \frac{e n}{B}  \int_0^\infty dE \left( -\frac{\partial f}{\partial E} \right) \frac{\tau_{\text{tr}} }{\left(1 + \omega_c^2 \tau_B^2(E)\right)\tau_B(E)},
\end{split}
\label{eq:sigmaxy}
\end{equation}
where $f(E) = [1 + \exp((E - E_F)/k_B T)]^{-1}$ is the Fermi-Dirac distribution.

Similar expressions for calculating magnetoresistance in the SdHO regime are presented in the review \cite{Dmitriev2012}, reffering these formulas to the approach developed in Refs.~\cite{Coleridge1989,Coleridge1996}. 
We emphasize that, instead of the conventional momentum relaxation time $\tau_{\text{tr}}$, the conductivity expressions contain the parameter $\tau_B$ introduced in Ref. ~\cite{Coleridge1989} as $\tau_{\text{tr}}$ renormalized by the ratio of the oscillating density of states in magnetic field to its zero-field value $\nu_0$:

\begin{equation}
\frac{1}{\tau_B} = \frac{1}{\tau_{\text{tr}}} \frac{\nu(E)}{\nu_0} .
\label{eq:tauB}
\end{equation}

This scaling facilitates the use of standard conductivity formulas for 2DEGs under SdHO conditions. It should be noted that the applicability of these expressions has been theoretically established within the SCBA framework ~\cite{Dmitriev2012, Dmitriev2003}, which is valid for $l_B > l_{\mathrm{corr}}$. In our experiment, the condition $l_B = l_{\mathrm{corr}}$ (with $l_{\mathrm{corr}} \approx 30$ nm) occurs at $B \approx 1$ T. Thus, for fields above $\sim$1 T, where high-amplitude SdHO are observed, the system is in the quasiclassical regime ($l_B < l_{\mathrm{corr}}$). A theoretical justification for the $\tau_B$ scaling in this particular regime is currently lacking. In the present work, we therefore examine the adequacy of such scaling as a phenomenological tool to describe the experimentally observed $R(B,T)$ dependencies across a wide magnetic field range that spans both SCBA and quasiclassical regimes.

\subsection{Landau Level Broadening}
To select the Landau level broadening model for analyzing high-amplitude SdHO with V-shaped minima, we focused on the magnetic field range of their observation (from 1~T to 3.5~T). For our samples in these fields, the conditions $l_B < l_{\mathrm{corr}} < R_c$ are satisfied, where $R_c$ is the cyclotron radius. The disorder correlation length was taken equal to the spacer thickness \cite{Dmitriev2012}, $l_{corr}=d = \SI{30}{\nano\meter}$. The magnetic length at $B = \SI{1}{\tesla}$ is \SI{26}{\nano\meter}, decreasing to \SI{14}{\nano\meter} at $B = \SI{3.5}{\tesla}$. The cyclotron radius at \SI{1}{\tesla} is \SI{140}{\nano\meter}, reducing to \SI{40}{\nano\meter} at \SI{3.5}{\tesla}.

The quasiclassical model \citep{Raikh1993} emerged as the most suitable theoretical framework for describing Landau level broadening in the majority of our experimental range ($B>1$ T), where $l_B$ becomes smaller than $l_{\mathrm{corr}}$. The applicability of alternative models and their comparison with experiment are presented in the following section on Dingle plot representation of SdHO. Although the applicability conditions for the quasiclassical approach \cite{Raikh1993} ($l_B \ll l_{\mathrm{corr}} \ll R_c$) are not strictly satisfied in our experiment, this model provides a better description of our measurements compared to alternatives, as demonstrated below.

To calculate the density of states, we used the Gaussian broadening of Landau levels, where the broadening width scales proportionally to the square root of the magnetic field \begin{equation*}
\Gamma = \sqrt{\frac{\hbar}{2 \pi \tauq}\hbar \omegac}.
\end{equation*} 
The density of states is expressed using the Poisson summation formula:

\begin{equation}
\begin{split}
    \nu(E) =  \nu_0 \left( 1 + 2 \sum_{s=1}^{\infty} (-1)^s \cos \left( 2\pi s\frac{E }{\hbar \omega_c} \right) exp {\left( -\frac{\pi s^2}{\omega_c \tau_q} \right)} \right),
\end{split}
\label{eq:pois}
\end{equation}
where $\nu_0 = m / (\pi \hbar^2)$ includes spin degeneracy. 

\begin{figure}[htbp]
\centering
\includegraphics[scale=1.5]{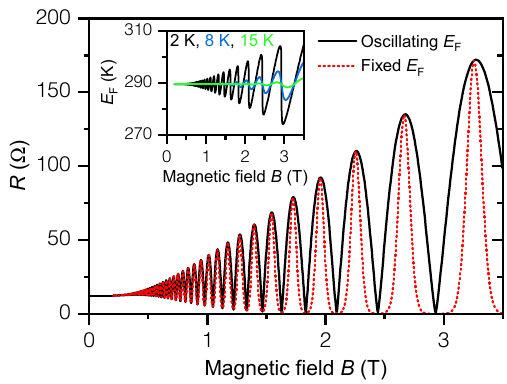}
\caption{
Theoretical magnetoresistance $R(B)$ at \SI{2}{\kelvin} without (dashed red) and with (solid black) $E_F$ oscillations. Inset: $E_F(B)$ oscillations at \SI{2}{\kelvin}, \SI{8}{\kelvin}, and \SI{15}{\kelvin}.
\label{fig2}
}
\end{figure}

Calculations with common assumption of fixed $E_F = n / \nu_0$ yield narrow resistance peaks and QHE-like zero plateaus (Fig.~\ref{fig2}, dashed red curve), that disagree with the experiment (see Fig.~\ref{fig1}a). This inconsistency demonstrates the need to account for $E_F$ oscillations with magnetic field.

\subsection{Fermi Level Oscillations}
We obtain the $B$-dependent Fermi energy $E_F(B)$ from expression for electron concentration: $n = \int_0^\infty \nu(E) f(E,E_F)  dE$. At \SI{2}{\kelvin}, $E_F$ exhibits almost abrupt jumps (see the inset in Fig.~\ref{fig2}). Increasing temperature suppresses these oscillations.

Magnetoresistance calculations incorporating the Fermi level oscillations yield broad peaks of oscillations with V-shaped minima (see Fig.~\ref{fig2}). Near even filling factors, these minima manifest as zero-resistance points with characteristic kinks rather than plateaus, matching experimental observations (Fig.~\ref{fig1}a). The observed kinks correspond to abrupt jumps of $E_F$ between adjacent Landau levels. This indicates that broadened Landau levels are separated by energy gaps without localized states where the Fermi level could pin.

We note that the necessity of accounting for the Fermi level oscillations under Landau quantization in fixed-density samples was previously highlighted in Ref.~\cite{Endo2008}. That work demonstrated that the magnetic-field dependence of the density of states near $E_F$ with magnetic-field-dependent Gaussian broadening produces a series of V-shaped oscillations featuring kinks at the minima. However, it did not provide magnetoresistance calculations.

\subsection{Fit to Data}

\begin{figure*}[htbp]
\centering
\includegraphics{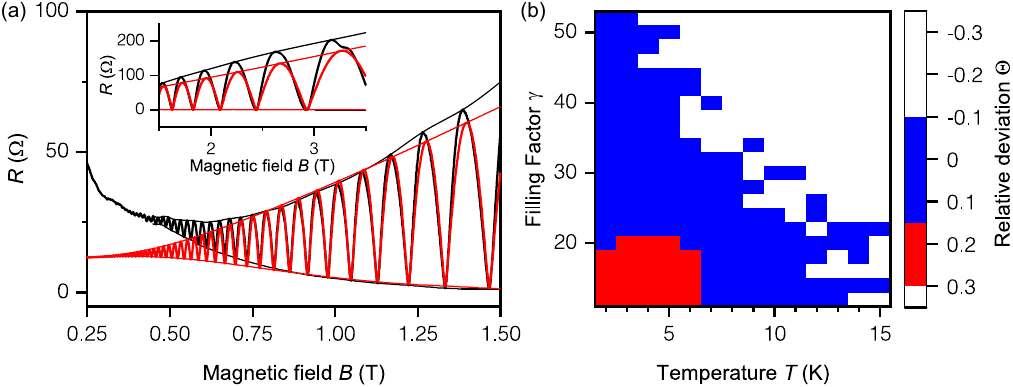}
\caption{
(a) Experimental and theoretical magnetoresistance $R(B)$ at \SI{2}{\kelvin} with upper/lower envelopes. Inset: High-field region showing deviation between the experiment and the theory due to spin splitting. (b) Color map of relative deviation of the envelope difference $\Theta$ versus filling factor $\gamma$ and temperature $T$.
\label{fig3}
}
\end{figure*}

We fit the experimental data with the theoretical magnetoresistance $R(B)$, calculated by Eqs.~\cref{eq:pois,eq:resistance,eq:sigmaxy,eq:sigmaxx,eq:tauB}, using $\tau_q$ and $\tau_{\text{tr}}$ as fitting parameters. Figure~\ref{fig3}a demonstrates the fitting results at $T = \SI{2}{\kelvin}$, with parameter values $\tau_q = \SI{2.15e-12}{\second}$ and $\tau_{\text{tr}} = \SI{8.4e-11}{\second}$. One can see excellent agreement between the theoretical model and the experiment in absolute resistance values across a wide magnetic field range. The obtained $\tau_{\text{tr}}$ value corresponds to $R_0 = \SI{12.2}{\ohm}$, an order of magnitude smaller than the measured zero-field resistance, modulated by ballistic effects (Fig. \ref{fig1}a). Notably, the obtained transport-to-quantum time ratio $\tau_{\text{tr}}/\tau_q \equiv N\approx$ 39, representing the number of small-angle scattering events required for momentum reversal, agrees well with the theoretical estimation $N \approx (k_F d)^2 \approx $ 39 for our samples and is consistent with Refs.~\cite{Dmitriev2003,Dmitriev2012}.

The discrepancy between the theoretical curve and the experimental data in the low-field region corresponds to the ballistic peak domain, while deviation above \SI{1.5}{\tesla} may arise from Zeeman splitting. Within the high-field range, experimental peaks at odd filling factors shift toward lower magnetic fields (see the inset in Fig.~\ref{fig3}a), and the peak above \SI{3}{\tesla} distinctly manifests spin splitting \cite{Leadley1998}.

A series of fitted $R(B)$ curves for temperatures from 2~K to 15~K is shown in Fig.~\ref{fig1}b. The qualitative agreement between the model and the experiment is clearly seen up to the highest temperatures. The model thus reproduces both the magnitude and the shape of oscillations, including their transformation with increasing temperature. The numerical fitting reveals that the quantum lifetime remains temperature-independent, while the resistance increases linearly with temperature from $R_0 = \SI{12.2}{\ohm}$ at $T = \SI{2}{\kelvin}$ to $R_0 = \SI{23.2}{\ohm}$ at $T = \SI{15}{\kelvin}$, indicating that $\tau_{\text{tr}}$ decreases by nearly a factor of two. Similar linear $R_0(T)$ behavior was previously observed in Ref.~\cite{Shi2014} and attributed to enhanced acoustic phonon scattering. Such scattering is not small-angle in nature; a single phonon scattering event can cause backscattering and significant increase of resistance without affecting the Landau level broadening.

To quantitatively compare the theory with the experiment across all ranges of magnetic field and temperature studied, we plotted upper $R_{\text{Upper}}(B)$ and lower $R_{\text{Lower}}(B)$ envelopes corresponding to resistance maxima and minima, respectively (see Fig.~\ref{fig3}a). The relative deviation 
\[
\Theta = \frac{(R_{\text{Upper}} - R_{\text{Lower}})_{\text{exp}} - (R_{\text{Upper}} - R_{\text{Lower}})_{\text{calc}}}{(R_{\text{Upper}} - R_{\text{Lower}})_{\text{exp}}}
\] 
between the experimental and calculated values is presented as a color map in Fig.~\ref{fig3}b as a function of the filling factor $\gamma$ and the temperature. The blue region indicates the theoretical-experiment deviation within 10\%. Quantitative agreement is observed for most filling factors and temperatures. However, at low filling factors and low temperatures, the experimental oscillation amplitude exceeds theoretical predictions (red region). These discrepancies likely arise from significant spin-splitting effects in strong magnetic fields \cite{Leadley1998,Piot2005}. Nevertheless, the maximal deviation between the experiment and the spin-independent model remains below 30\%. At temperatures above \SI{8}{\kelvin}, where spin-splitting effects are suppressed, excellent agreement is restored even at low filling factors.

Thus, our model based on the Gaussian-broadened Landau levels density of states \eqref{eq:pois} and the $B$-scaling of the scattering time \eqref{eq:tauB} accurately describes the observed SdHO over the magnetic field range exceeding the applicability limit of the model. The entire experimental dataset is described with a single $\tau_q$ independent of both $T$ and $B$, while $\tau_{\text{tr}}$ decreases inversely with temperature due to the acoustic phonon scattering. We note that although the assumption about $E_F$ oscillations enables reproduction of V-shaped oscillations, the oscillation amplitude (half the envelope difference) itself is independent of $E_F$ variations. This occurs because envelopes are constructed at $R(B)$ points corresponding to integer filling factors where the oscillating $E_F$ coincides with its constant value counterpart (see Fig.~\ref{fig2}).

It is important to discuss the applicability limits of our model, which is grounded in the Kubo formalism and relies on the concepts of transport and quantum scattering times. In  2DEG at high magnetic fields electron motion can become non-ergodic, and transport may transition from a diffusive to a percolative regime, where electrons drift along closed equipotential contours of a smooth random potential \cite{Fogler1997}. In such a regime, the diffusive approach and the notion of a scattering time become questionable. Although the condition $l_{\mathrm{corr}} \gg \lambda_F$, where $\lambda_F$ is the Fermi wavelength, is not strictly met in our samples ($l_{\mathrm{corr}}$ is comparable to $\lambda_F$, 29 nm in our samples) — a condition explicitly stated as necessary in \cite{Fogler1997} — several key observations also indicate that our high-mobility 2DEG with weak disorder remains within the diffusive/ergodic regime even at high magnetic fields. First, the observed V-shaped resistance minima with zero-resistance points, instead of quantum Hall plateaus, provide direct evidence for the absence of a significant density of localized states in the Landau level gaps, which would otherwise pin the Fermi level. Second, the experimentally measured non-local resistance in the SdHO regime is independent of the specific realization of the random potential and remains negligible compared to the longitudinal resistance. Finally, the model's quantitative accuracy across Hall bars differing in size by an order of magnitude (see Sec. VI) suggests scale-invariant, self-averaging transport properties, as expected for a diffusive system. Therefore, we conclude that the use of a local conductivity tensor is justified for describing our experimental data.

\section{Dingle Plot Analysis and Model Comparison}
\label{sec:dingle}

Dingle plot analysis is based on the fact that in weak magnetic fields ($\omega_c \tau_q \ll 1$), the oscillations of the density of states are described by the first harmonic \cite{Coleridge1989} in \eqref{eq:pois}:
\begin{equation*}
\Delta g = -2\nu_0 \exp\left(-\frac{\pi}{\omegac \tauq}\right)  \cos\left(2\pi \frac{\EF}{\hbar \omegac}\right) .
\label{eq:first}
\end{equation*}

This expression holds for both the Lorentzian broadening model \cite{Coleridge1989} and the modified SCBA model \cite{Vavilov2004} derived within the SCBA framework, as well as for the magnetic-field-dependent Gaussian model obtained via the quasiclassical approach. For all these models, the following relation is valid:

\begin{equation}
\frac{\Delta R}{4 R_0 D_T(2\pi^2 k T / \hbar \omega_c)} = exp{\left(-\frac{\pi}{\omega_c \tau_q}\right)}, 
\label{eq:Dingle}
\end{equation}
where $D_T(x) = x / \sinh x$.

\begin{figure}[htbp]
\centering
\includegraphics[scale=3.5]{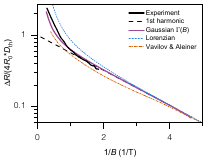}
\caption{
Dingle plot: Experimental (black solid) and theoretical Dingle factors versus $1/B$ for Gaussian (red solid), Lorentzian (blue fine dashed), and the modified SCBA \cite{Vavilov2004} (green dash-dotted) broadening models. Thick dashed line: First harmonic approximation. All data is shown for $T=$ 4~K.
\label{fig4}
}
\end{figure}

The right-hand side of Eq.~\eqref{eq:Dingle} represents the so-called Dingle factor $\delta$. This quantity is plotted logarithmically versus $1/B$ in Fig.~\ref{fig4} for both the experimental $R(B)$ data and the theoretical dependencies at $T=$ 2~K: our magnetic-field-dependent Gaussian broadening model, the Lorentzian broadening model, and the modified SCBA model \cite{Vavilov2004} (see Eq.~(4.18) in Ref.~\cite{Vavilov2004}). The linear thick dashed curve corresponds to the first harmonic approximation using the parameters $\tau_{\text{tr}}$ and $\tau_q$ obtained from fitting in the previous section. Here, $\Delta R$ for all models and experimental data was defined as half the envelope difference: $\Delta R = [R_{\text{Upper}}(B) - R_{\text{Lower}}(B)]/2$, corresponding to the SdHO amplitude at small oscillations.

At large $1/B \approx \SI{4}{\per\tesla}$ ($B \approx \SI{0.25}{\tesla}$), all models exhibit identical linear behavior consistent with the first harmonic. As $1/B$ decreases ($B > \SI{0.25}{\tesla}$), both SCBA models deviate significantly from the straight line of first harmonic (Fig.~\ref{fig4}), indicating substantial contributions from higher-order harmonics to magnetoresistance oscillations. However, at lower magnetic fields ($B < \SI{0.5}{\tesla}$), SdHO remain experimentally unobservable in our measurements, and magnetoresistance is strongly distorted by ballistic effects. Consequently, these models are unsuitable for the extraction of $\tau_{\text{tr}}$ and $\tau_q$ from the experimental data.

We note that the magnetic-field-dependent Gaussian model maintains agreement with the first harmonic over a wider $1/B$ range. The experimental dependence follows the first harmonic only within a narrow interval ($\SI{1}{\per\tesla} < 1/B < \SI{1.5}{\per\tesla}$). Nevertheless, both experimental and calculated Gaussian curves show close correspondence, while other models exhibit substantial deviations from our data.

Dingle plot analysis further demonstrates that the magnetic-field-dependent Gaussian Landau level broadening model accurately describes SdHO behavior even at low magnetic fields ($\SI{0.5}{\tesla} < B < \SI{1}{\tesla}$) - the transitional regime between quasiclassical and SCBA approaches.

In our experiments, even at $\SI{2}{\kelvin}$, SdHO remain suppressed below $\SI{0.5}{\tesla}$, with $R_0$ obscured by the ballistic peak. Therefore, direct comparison of experimental magnetoresistance with theoretical models via standard Dingle plot analysis is unfeasible.

\section{Experiment in a Macroscopic Hall bar}

\begin{figure}[htbp]
\centering
\includegraphics[scale=1.5]{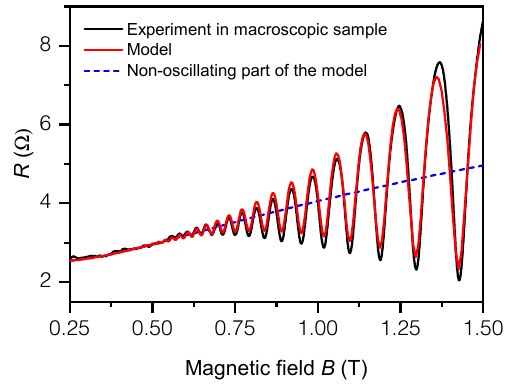}
\caption{
Magnetoresistance $R(B)$ for a macroscopic Hall bar ($15\times10$~\si{\micro\meter\squared}) at \SI{4}{\kelvin}. Black solid: experiment. Red solid: full model. Dashed blue: nonoscillating background from Eq.~\eqref{eq:nonosc}, accounting for thermal suppression of SdHO.
\label{fig5}
}
\end{figure}

As demonstrated above, our model enables determination of $R_0$ from high-amplitude SdHO in strong magnetic fields for microscopic Hall bars where ballistic effects at low fields obscure $R_0$. To extend the model's applicability to lower magnetic fields and cross-check $\tau_{\text{tr}}$ and $R_0$ values for micro/nanostructures, we measured $R(B)$ in a macroscopic Hall bar ($W = \SI{15}{\micro\meter}$, $L = \SI{10}{\micro\meter}$) fabricated from the same 2DEG as the microscopic Hall bars. Measurements were performed at $T = \SI{4}{\kelvin}$ (Fig.~\ref{fig5}). The magnetoresistance of this sample lacks a significant ballistic peak at low fields, instead exhibiting smooth positive magnetoresistance up to $B = \SI{0.6}{\tesla}$. We fitted the experimental $R(B)$ data using our model in the range from 0.75~T to 1.5~T where high-amplitude SdHO are observed, finding excellent agreement extending to low fields where SdHO are thermally suppressed and the magnetoresistance increases smoothly.

The observed smooth positive magnetoresistance originates from density of states modulation. Transport-active electrons predominantly occupy high-density-of-states regions, causing the effective density of states $\nu_{\mathrm{eff}}$ to systematically exceed its zero-field value $\nu_0$. The $\nu_{\mathrm{eff}}$ magnitude  grows monotonically with $B$ as the reduced Landau level overlap narrows density-of-states peaks, enhancing the contribution of the local maxima of density of states to the average value. Consequently, increasing $\nu_{\mathrm{eff}}$ with the magnetic field increases $\sigma_{xx}$, yielding the observed positive magnetoresistance.

Notably, under the condition $\omega_c \tau_{\text{tr}} \gg 1$ (satisfied for our samples at $B > \SI{0.1}{\tesla}$), the unity term in the denominators of the conductivity tensor components (Eqs.~\eqref{eq:sigmaxx} - \eqref{eq:sigmaxy}) becomes negligible. Substituting $\tau_B$ from Eq. \eqref{eq:tauB} into the resulting expression and neglecting all terms except $\left(\frac{e n}{B}\right)^{2}$ in the denominator yields:

\begin{equation}
R = R_0  \int_0^\infty \left(\frac{\nu(E)}{\nu_0} \right)^2 \left( -\frac{\partial f}{\partial E} \right) dE.
\end{equation}

Substituting the density of states expression from Eq.~\eqref{eq:pois} yields:
\begin{equation*}
\begin{split}
R = R_0 \int_{0}^{\infty} \left[ 1 + 2 \sum_{s=1}^{\infty} (-1)^s \cos \left( 2 \pi s \frac{E}{\hbar \omega_c} \right) \delta^{s^p} \right]^2 \left( -\frac{\partial f}{\partial E} \right) dE
\end{split}
\end{equation*}
where $p = 1$ for Lorentzian broadening and $p = 2$ for Gaussian broadening.

The resistance decomposes into two parts:
\begin{equation*}
R = R_{\mathrm{nonosc}} + R_{\mathrm{osc}}
\end{equation*}
For Gaussian broadening, the nonoscillating part is given by:
\begin{equation}
\begin{split}
    \frac{R_{\mathrm{nonosc}}}{R_0} =  1 + 2 \sum_{s=1}^{\infty} (\delta^2)^{s^2} =  \theta_3(0, \delta^2)
\end{split}
\label{eq:nonosc}
\end{equation}
where $\theta_3$ denotes the Jacobi theta function. This expression accurately describes the magnetoresistance behavior under thermally suppressed SdHO conditions. For visual reference, the dashed curve in Fig.~\ref{fig5} shows the $R(B)$ dependence calculated using Eq.~\eqref{eq:nonosc}.

The emergence of a positive magnetoresistance background within our model, despite its roots in the Drude formalism, deserves clarification. The key point is that the substitution $1/\tau_B \propto \nu(E)$ injects a quantum-mechanical ingredient -- the oscillating density of states -- into the classical transport framework. In the high-temperature regime where SdHO are suppressed ($D_T(x) \to 0$), the resistance in Eq.~\eqref{eq:nonosc} is determined by the average of the squared density of states, $\langle[\nu(E,B)]^{2}\rangle$. It should be noted, that similar proportionality was observed in \cite{Mayer2016, Dmitriev2005}. While the average density of states $\langle\nu(E,B)\rangle$ remains constant and equal to $\nu_0$, the variance $\langle[\Delta\nu(E,B)]^{2}\rangle$ grows with magnetic field as the Landau levels narrow and their overlap decreases. Since $\langle[\nu]^{2}\rangle = \nu_0^{2} + \langle[\Delta\nu]^{2}\rangle$, this leads to a monotonic increase of resistance with $B$, explaining the smooth positive background. From a classical viewpoint, this effect could be interpreted as resulting from a dispersion in carrier properties: electrons at different energies across the broadened Fermi level experience different local densities of states, effectively acting as multiple carrier groups, which is a known mechanism for classical positive magnetoresistance. An alternative classical interpretation might involve memory effects in transport through a disordered potential \cite{Dmitriev2012}. A rigorous derivation of this effect from purely classical principles in quantizing magnetic fields remains a subject for future theoretical work.

The smooth positive magnetoresistance background observed here and in other high-mobility 2DEGs \cite{Shi2014} should be distinguished from the positive magnetoresistance reported in systems exhibiting a transition to non-ergodic, percolative transport at high magnetic fields \citep{Floser2013}. In the latter case, the positive magnetoresistance is observed in systems where well-developed quantum Hall plateaus at low temperatures signify a substantial density of localized states. The positive magnetoresistance in that regime is interpreted in \citep{Floser2013} as arising from activation from these localized states associated with classical drift along closed equipotential contours of a smooth disorder potential. In contrast, our samples show deep V-shaped minima with zero resistance points instead of quantum Hall plateaus, indicating a negligible density of localized states. The positive magnetoresistance in our case finds a natural explanation within the diffusive paradigm, as a direct consequence of the increasing density of states at the Fermi level due to reduced Landau level overlap, as described by Eq. ~\eqref{eq:nonosc}.

The oscillating resistance components for different Landau level broadening types can be compared separately. Considering the first few harmonics, for Gaussian broadening:

\begin{equation*}
\begin{split}
 \frac{R_{\mathrm{osc}}^{\mathrm{G}}}{R_0} = 
-4\delta \cos \left( 2\pi \frac{E_F}{\hbar \omega_c} \right) D_T \left( \frac{2\pi^2 kT}{\hbar \omega_c} \right) 
+ \\ + 2\delta^2 \cos \left( 2\cdot 2\pi \frac{E_F}{\hbar \omega_c} \right) D_T \left( 2\cdot \frac{2\pi^2 kT}{\hbar \omega_c} \right) 
+ \mathcal{O} \left( \delta^4 \right)   
\end{split}
\end{equation*}

while for Lorentzian broadening:

\begin{equation*}
\begin{split}
 \frac{R_{\mathrm{osc}}^{\mathrm{L}}}{R_0} = 
-4\delta \cos \left( 2\pi \frac{E_F}{\hbar \omega_c} \right) D_T \left( \frac{2\pi^2 kT}{\hbar \omega_c} \right) 
+ \\ + 6\delta^2 \cos \left( 2\cdot 2\pi \frac{E_F}{\hbar \omega_c} \right) D_T \left( 2\cdot \frac{2\pi^2 kT}{\hbar \omega_c} \right) 
+ \mathcal{O} \left( \delta^3 \right)   
\end{split}
\end{equation*}

The first harmonics coincide completely for both broadening types, while the second harmonic amplitude for Lorentzian broadening is three times larger than for Gaussian. Thus, constructing the first few harmonics is often sufficient to compare with experimental magnetoresistance and identify the broadening type realized experimentally.

Our two-parameter model, reproducing giant SdHO with V-shaped minima in high quantizing magnetic fields, also accurately describes the smooth positive magnetoresistance of macroscopic 2DEG Hall bars across the entire temperature range (2--15 K). Although the model does not explicitly incorporate a specific phonon scattering mechanism, the temperature dependence of the extracted transport scattering time $\tau_{\mathrm{tr}}(T)$ provides valuable insight. We find that the zero-field resistance $R_0 \propto 1/\tau_{\mathrm{tr}}$ increases linearly with temperature. This linear increase is a hallmark of scattering by acoustic phonons in this temperature regime. Consequently, the extracted $\tau_{\mathrm{tr}}$ itself decreases with temperature in a manner consistent with this mechanism. In contrast, the quantum scattering time $\tau_q$ remains temperature-independent. The fact that our phenomenological model --- with its sole temperature dependence entering via $\tau_{\mathrm{tr}}(T)$ --- successfully reproduces both the giant SdHO and the smooth positive background at all temperatures strongly suggests that it correctly captures the dominant effect of phonon scattering on the magnetoresistance. This result confirms the validity of $\tau_{\mathrm{tr}}$ extraction via our method, particularly for samples where a large ballistic peak precludes the direct determination of $R_0$.

\section{Conclusion}

We experimentally studied giant SdHO exhibiting V-shaped minima in a high-mobility 2DEG with weak disorder potential. The phenomenological model with two fitting parameters $\tau_{\text{tr}}$ and $\tau_q$, developed here, provides accurate description of the experimental data across wide magnetic field (from 0~T to 3.5~T) and temperature (from 2~K to 15~K). This model reproduces not only the oscillatory form of magnetoresistance but also the absolute oscillation amplitude, including deep V-shaped minima with zero-resistance points in quantizing fields. It further describes smooth positive magnetoresistance in regions where SdHO are thermally suppressed. Gaussian broadening produces exponentially sharp density-of-states decay at Landau level edges, causing abrupt Fermi level jumps between adjacent levels at even filling factors.

Experimental observation of zero-resistance $R(B)$ minima and their precise theoretical reproduction provide direct evidence of negligibly low background density of localized states within Landau level energy gaps. This is achievable only in ultra-high-mobility samples with minimal disorder. Weak disorder potential in our structures was realized by incorporated X-valley electron superlattices  that effectively screen Coulomb potentials of dopant impurities. Additionally, dual-side doping enhances $\tau_{\text{tr}}$. The high electron density further reduces disorder impact through improved screening.

Our theoretical model enabled extraction of temperature-dependent 2DEG parameters: $\tau_{\text{tr}}$ decreases inversely proportional to the temperature from \SI{2}{K} to \SI{15}{K}, attributed to activated acoustic phonon scattering that efficiently induces backscattering. Meanwhile, $\tau_q$ remains temperature-independent, indicating dominant static disorder scattering. Crucially, $\tau_{\text{tr}}$ was reliably determined even in microstructures where ballistic effects distort intrinsic 2DEG resistance by orders of magnitude.

We propose this model as a powerful tool for analyzing magnetoresistance across wide magnetic field ranges and performing 2DEG parameter spectroscopy, particularly valuable for micro/nanostructures with high-mobility 2DEGs and weak disorder potentials.

\section*{Acknowledgments}
This work is supported by the Russian Science Foundation
(Grant No. 22-12-00343—$\Pi$ experimental study) and the Ministry of
Science and Higher Education of the Russian Federation (Project No.
FWGW-2025-0023 initial heterostructures characterization).

\end{document}